\documentclass[aps,pra,twocolumn,superscriptaddress, amsmath,amssymb,10pt]{revtex4-2}
\usepackage[colorlinks=true,linkcolor=blue,urlcolor=blue,citecolor=blue,pdfusetitle]{hyperref}
\usepackage{graphicx}
\usepackage{color}
\usepackage{orcidlink}

\begin{document}
\title{Stable collective charging of ultracold atoms quantum batteries}
\author{Abel Rojo-Francàs~\orcidlink{0000-0002-0567-7139}}
\email{abel.rojo@fqa.ub.edu}
\affiliation{Departament de F{\'i}sica Qu{\`a}ntica i Astrof{\'i}sica, Facultat de F{\'i}sica, Universitat de Barcelona, E-08028 Barcelona, Spain}
\affiliation{Institut de Ci{\`e}ncies del Cosmos, Universitat de Barcelona, ICCUB, Mart{\'i} i Franqu{\`e}s 1, E-08028 Barcelona, Spain.}

\author{Felipe Isaule~\orcidlink{0000-0003-1810-0707}}
\email{felipe.isaule@uc.cl}
\affiliation{Instituto de Física, Pontificia
Universidad Católica de Chile,
Avenida Vicuña Mackenna 4860,
Santiago, Chile}

\author{Alan C. Santos~\orcidlink{0000-0002-6989-7958}}
\email{ac\_santos@iff.csic.es}
\affiliation{Instituto de Física Fundamental (IFF), Consejo Superior de Investigaciones Científicas (CSIC), Calle Serrano 113b, 28006 Madrid, Spain}

\author{Bruno Juliá-Díaz~\orcidlink{0000-0002-0145-6734}}
\email{bruno@fqa.ub.edu}
\affiliation{Departament de F{\'i}sica Qu{\`a}ntica i Astrof{\'i}sica, Facultat de F{\'i}sica, Universitat de Barcelona, E-08028 Barcelona, Spain}
\affiliation{Institut de Ci{\`e}ncies del Cosmos, Universitat de Barcelona, ICCUB, Mart{\'i} i Franqu{\`e}s 1, E-08028 Barcelona, Spain.}

\author{Nikolaj Thomas Zinner~\orcidlink{0000-0002-5972-9790}}
\email{zinner@phys.au.dk}
\affiliation{Department of Physics and Astronomy, Aarhus University, DK-8000 Aarhus C, Denmark}

\begin{abstract}
We propose a novel quantum battery realized with a few interacting particles in a three-well system with different on-site energies, which could be realized with ultracold atom platforms. We prepare the initial state in the lowest energy well and charge the battery using a Spatial Adiabatic Passage 
(SAP)-based protocol, enabling the population of a higher energy well. We examine the charging under varying interaction strengths and reveal that the consideration of collective charging results in an intriguing oscillatory behavior of the final charge for finite interactions, through diabatic evolution. Our findings open a new avenue for building stable and controllable quantum batteries.

\end{abstract}

\maketitle

\section{introduction}
Energy storing quantum devices~\cite{alicki_entanglement_2013,binder_quantacell_2015} emerged as part of a quantum energy initiative for quantum-inspired technologies~\cite{auffeves_quantum_2022}, where the energy stored in these \textit{quantum batteries} (QBs) would be used for further transfer to quantum consumption hubs. The performance of QBs is mainly ruled by laws of quantum thermodynamics, which dictates physical processes involving entropy production, heat, and work in the quantum realm~\cite{deffner_quantum_2019}. Through this theory, one may properly manage useful work provided by quantum systems at single atom level~\cite{rosnagel_single-atom_2016}, the energy cost to perform quantum tasks~\cite{monsel_energetic_2020,deffner_energetic_2021}, and to predict the extractable amount of work stored in such systems~\cite{allahverdyan_maximal_2004}, among other applications~\cite{myers_quantum_2022}. By harnessing genuine effects from interacting quantum systems, QBs exhibit scalable enhanced charging performance (power) with respect to their non-interacting (classical) counterpart~\cite{barra_dissipative_2019,andolina_extractable_2019,ferraro_high-power_2018,campaioli_enhancing_2017,crescente_ultrafast_2020,carrega_dissipative_2020,gyhm_beneficial_2024,gyhm_quantum_2022,kim_operator_2022,rossini_quantum_2020,rosa_ultra-stable_2020,shaghaghi_micromasers_2022,rodriguez_artificial_2023}. Experimental realizations of QBs have been done with superconducting integrated circuits~\cite{hu_optimal_2022,gemme_ibm_2022}, Nuclear Magnetic Resonance~\cite{joshi_experimental_2022}, quantum dots~\cite{maillette_de_buy_wenniger_experimental_2023}, and organic microcavities~\cite{quach_superabsorption_2022}, with the first experiment of a room-temperature QB designed with carboxylate-based metal complexes~\cite{cruz_quantum_2022}.

Despite the high performance of these devices, some properties of such devices make their usage for real-world applications a challenge. In particular, and the main focus of this work, \textit{instantaneous discharging} is observed for always-on charging of QBs due to the quantum recurrence theorem of Poincaré~\cite{bocchieri_quantum_1957}, demanding then for the development of stable charging strategies. The instantaneous discharging is related to oscillations in time of the stored energy in the battery, leading to loss of performance due to the energy backflow from the battery to the charger~\cite{moraes_charging_2022}. Proposals of stable charging of QBs have been induced through localization effects in disordered spin systems~\cite{arjmandi_localization_2023,rossini_many-body_2019}, using single~\cite{santos_stable_2019} and collective dark-states~\cite{quach_using_2020}, Zeno protection~\cite{gherardini_stabilizing_2020}, adiabatic evolutions~\cite{santos_stable_2019,santos_stable_2020}, nonreciprocal approach~\cite{ahmadi_nonreciprocal_2024} and transitionless driving~\cite{moraes_charging_2022}, for example, with the first experimental implementation of optimal stable charging done with superconducting three-level atoms~\cite{hu_optimal_2022}.
\begin{figure}[t]
	\includegraphics[width=\columnwidth]{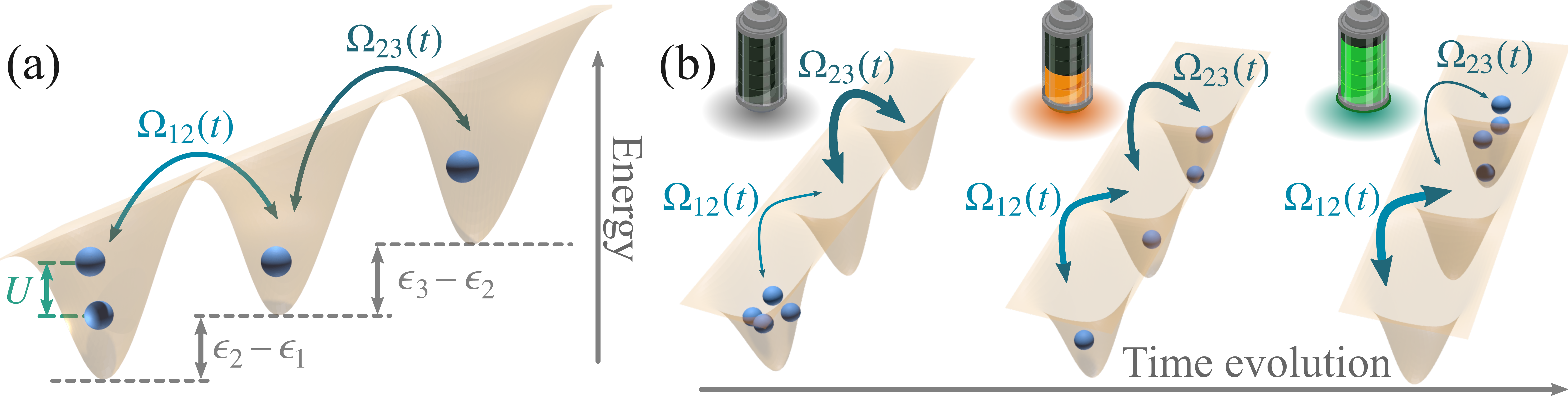}
	\caption{(a) The physical system consisting of $N$ particles which may populate three wells with different energy. (b) Schematic representation of the evolution used for the stable charging protocol, in which the tunneling strengths $\Omega_{12}(t)$ and $\Omega_{23}(t)$ are switched on in a counter-intuitive way to avoid populating the intermediate well with energy $\epsilon_{2}$.} \label{fig:scheme}
\end{figure}
Inspired by QBs proposals with interacting spin chains~\cite{le_spin-chain_2018,ghosh_enhancement_2020,zhao_quantum_2021,ghosh_dimensional_2022},  ultracold atoms confined in optical lattices appear as a promising platform for designing QBs~\cite{konar_quantum_2022}. The high level of control offered by ultracold atoms~\cite{bloch_quantum_2012,gross_quantum_2017,lewenstein_ultracold_2012,lewenstein_ultracold_2007b} 
makes them unique candidates for developing efficient and stable QBs. In particular, the control over the atomic internal states and of the inter-atomic interactions offer novel configurations of QBs not achievable with other quantum physical systems.
\begin{figure*}[t]
	\centering
	\includegraphics[width=1\textwidth]{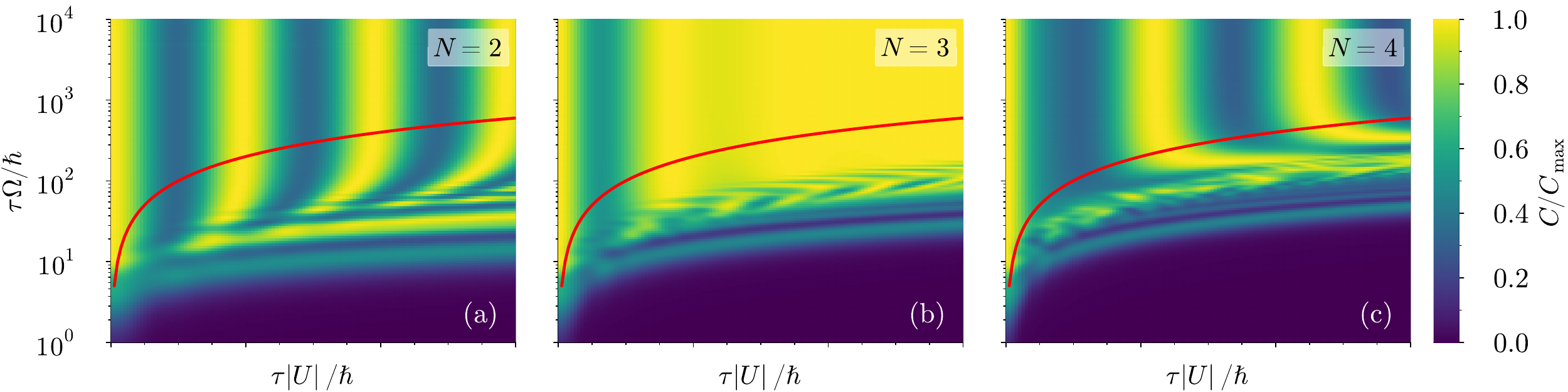}
	\caption{Normalized final ($t=\tau$) charge $C/C_{\mathrm{max}}$ as a function of the interaction $\tau |U|/\hbar$ and the coupling $\tau \Omega/\hbar$, for $N=2$, $3$, and $4$, as indicated in the legends. The solid red line on the charge map indicates the line with $|U|=0.1\Omega$.}
	\label{fig:charge_power}
\end{figure*}

In this article, we propose a three-well QB controlled with an SAP-based protocol~\cite{bradly_coherent_2012,menchon-enrich_spatial_2016,rubio_optimal_2016}, as depicted in Fig.~\ref{fig:scheme}a. The SAP implementation is already possible in ultracold atoms laboratories~\cite{taie_spatial_2020}. Moreover, and in contrast to previous related studies~\cite{santos_stable_2019}, we consider quantum collective effects by including on-site interactions. The interaction effects cause the time evolution to no longer be adiabatic. However, since we are using the SAP order of applying the couplings and the final target state, we refer to our setup as an SAP-based protocol. Our protocol creates a stable system if we achieve the maximum charge, and in this study, we will focus on the charging mechanism. In addition to the proposed three-well implementation, such a QB could also be realized with an array of three-level systems with a Stimulated Raman Adiabatic Passage-based protocol~\cite{vitanov_stimulated_2017,bergmann_roadmap_2019}.

This article is organized as follows. In Sec.~\ref{sec:model} we present the model we are working with, including the Hamiltonian and the charge definition.
Next, in Sec.~\ref{sec:int.charge} we explore how the interaction affects the charging protocol. 
In Sec.~\ref{sec:analytical} we use analytical models to explain the effects of the interaction on the charge, being exact for the two-particle case and an approximation for a larger number of particles. 
Finally, in Sec.~\ref{sec:conclusions} we summarize our work and we present the conclusions.

\section{Model}\label{sec:model}
We will consider as a general case, $N$ identical bosons in a three-well system, where the 
wells energy is $\epsilon_i$, with $i=1$ to $3$ in ascending order. In addition, we will consider an energy $U$ 
when a pair of particles are in the same well.
On top of that, our SAP-inspired protocol considers a time-dependent tunneling
between the wells $1$ and $2$ and also, between the wells $2$ and $3$. As illustrated in Fig.~\ref{fig:scheme}, the Hamiltonian in a second quantization formalism reads as $\hat{H}=\hat{H}_0+\hat{H}_\mathrm{coll}+\hat{H}_{\mathrm{ch}}$, with $\hat{H}_0=\sum_{i=1}^3 \epsilon_i \hat{b}^\dagger_i \hat{b}_i$ the self-Hamiltonian of the system that sets the energy scale of the battery, $\hat{H}_\mathrm{coll}=\frac{1}{2}U\sum_{i=1}^3 \hat{n}_i(\hat{n}_i-1)$ is the in-situ particle-particle interaction that describes the collective (quantum) aspect of the battery charging, and the charging $\hat{H}_{\mathrm{ch}}$ describes the time-dependent tunneling between wells
\begin{equation}
\hat{H}_{\mathrm{ch}}=\Omega_{12}(t)e^{i\omega_{12}t} \hat{b}_1^\dagger \hat{b}_2 +\Omega_{23}(t) e^{i\omega_{23}t} \hat{b}_2^\dagger \hat{b}_3 +\mathrm{h.c.}\,,
\end{equation}
where $\hat{b}^\dagger_i$ ($\hat{b}_i$) is the creation (annihilation) operator of 
a particle in the well $i$, and $\hat{n}_i=\hat{b}^\dagger_i\hat{b}_i$ is the number operator. The coupling is assumed to have driving strength $\Omega_{ij}(t)$ in resonance with the energy transitions of the system at a frequency 
$\omega_{ij}=\left(\epsilon_j-\epsilon_i\right)/\hbar$. 
The SAP is implemented by the suitable choice of the tunneling such that $\Omega_{12}(0)=\Omega_{23}(\tau)=0$ and $\Omega_{12}(\tau), \Omega_{23}(0)\neq 0$. Here we choose a lineal transition given by $\Omega_{12}(t)= t\Omega/\tau$ and $\Omega_{23}(t) = \Omega - \Omega_{12}(t)$. So, according to the adiabatic theorem~\cite{kato_adiabatic_1950}, for large enough $\tau$  the system will move the particles from the lowest to the highest energy well.

At the end of the charging, the extractable amount of energy by coherent drives from a QB is defined by erogotropy~\cite{allahverdyan_maximal_2004}, which for a unitary evolution charging can be uniquely obtained as~\cite{santos_stable_2020}
\begin{equation}
  C=\langle\Psi(\tau)|\hat{H}_0|\Psi(\tau)\rangle\,,
\end{equation}
where $|\Psi(\tau)\rangle$ is the 
state at the end of the SAP protocol. By exploiting the definition of battery capacity as the energy of the most active state~\cite{yang_battery_2023}, it is possible to get the maximum charge for our system with $N$ 
particles as $C_\mathrm{max.}=N\epsilon_3$, corresponding to an eigenstate of the final Hamiltonian and thus, stable. The energy charging of the QB may be obtained from the Schrödinger equation in the interaction picture as
\begin{equation}
    \label{eq:Hamiltonian_rot.frame}
    \hat{H}_\mathrm{int}=\hat{H}_\mathrm{coll}+\hbar \Omega_{12}(t)\hat{b}^\dagger_1\hat{b}_2 + \hbar\Omega_{23}(t)\hat{b}^\dagger_2\hat{b}_3 +\mathrm{h.c.} ,
\end{equation}
and thus the charging performance depends only on the tunneling amplitudes and the collectivity parameter $U$.

\section{Interacting system charge}\label{sec:int.charge}
Now, we will analyze the final charge diagram 
as a function of the coupling $\tau\Omega$ and the internal interaction $\tau U$. In Fig.~\ref{fig:charge_power} we show the charge diagram corresponding to 
$N=2$ to $N=4$.
The charge only depends on the absolute value of the interaction $|U|$, but not on its sign.
The energy spectra are computed with negative interactions  $U<0$ for description simplicity, but the physics is the same for $U>0$.

Firstly, we note that the non-interacting limit $U=0$ (left side of the panels) corresponds to the single-particle configuration studied with a STIRAP protocol in
Ref.~\cite{santos_stable_2019}, which does not consider collective charging. For $|U|>0$, Fig.~\ref{fig:charge_power}  shows two distinct regions. On one side, when the coupling $\tau\Omega$ is much smaller than the 
interaction $\tau U$ (bottom right regions of the panels), the system ends
with essentially a null final charge ($C\approx 0$). On the other hand, when the coupling is much larger than the interaction (top regions of the panels), the system produces a finite final charge $C$ but also shows an intriguing oscillation of $C$ with increasing $\tau|U|$ between maximum and partial charging.
\begin{figure}[t]
    \centering
    \includegraphics[width=1\columnwidth]{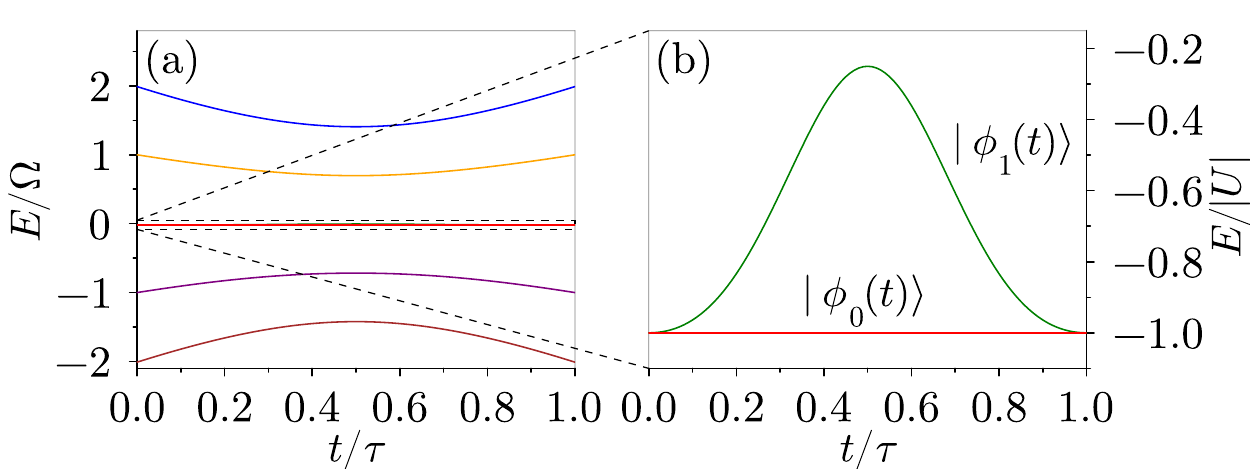}
    \caption{Energy spectrum for two particles with interaction $U=-0.02\Omega$. Panel (a) shows the complete energy spectrum. Panel (b) zooms the two states in the central manifold. Note that panels (a) and (b) plot $E/\Omega$ and panel $E/U$, respectively. }
    \label{fig:spectrum_2}
\end{figure}

To understand why the system does not charge in the weak tunneling region, we examine the system with $N=2$, even though our arguments can be generalized for more particles.
A system with two bosons on a triple well potential has six states 
in its Hilbert space. Due to the large interaction $U$, the system will form two manifolds,
one with the three states with the particles in the same well and the other
with both particles in different wells. Both manifolds are 
separated by an energy gap $\sim U$. In addition, our tunneling is a single-particle operator and thus it can only 
move the particles individually. The latter connects the states with two particles 
in the same well to states with particles at different wells. If the 
coupling is weak compared with the gap between the states that are connecting, the 
transition is almost suppressed. Therefore, by starting from both particles 
in the lowest energy well, if $U\gg \Omega$ the state cannot evolve with time, and thus both particles remain in the same initial site during the evolution, preventing charging.

In the second region, where the coupling is much larger than the interaction (above the solid red line in the panels), the results do not depend on the
value of the coupling $\tau\Omega$. We find oscillations
in the charge as a function of $|U|$ for any number of particles $N$, where the period of these oscillations depends on $N$.
Interestingly, we also find that systems with odd and even numbers of particles show a distinct onset of oscillations with $U$. Indeed, with even $N$ the oscillations are approximately 
periodic and with a fixed amplitude. In contrast, for odd $N$ the amplitude of the
oscillations decreases but achieves periodic full charging. We will explain this phenomenon
later on.

Importantly, our results show that it is possible to reach $C_\mathrm{max}$ for any $N$ and if one chooses $|U|$ to the regions of maximum charge. This feature could be exploited in experiments to get a full charge by tuning the inter-atomic interactions. The maximum charge is also achievable for a finite interaction when the interaction strength and the coupling have similar values. That situation is interesting, but we will focus on the situation where $\Omega\gg|U|$. 

\section{Analytical time-evolution}\label{sec:analytical}
Having examined the charging from our numerical calculations, we now provide an analytical interpretation of our results. We will focus on the case with $\Omega/|U| \gg 1$, corresponding to the region above the solid red lines in Fig.~\ref{fig:charge_power}(a-c). 

We first analyze the two-particle case, which can be examined analytically for $\Omega/|U| \gg 1$.
By taking $t$ as a fixed parameter in Eq.~(\ref{eq:Hamiltonian_rot.frame}), we can diagonalize the Hamiltonian for two particles analytically, resulting in an energy spectrum as the one shown in Fig.~\ref{fig:spectrum_2}. The spectrum has states
with an energy proportional to $\pm\Omega$ and others to $\pm2\Omega$. While these states show a weak dependence on $U$ and $t/\tau$, the gap between these states is $\sim \Omega$.
In addition, at the center of the panel (a) the spectrum shows two states whose energy does not depend
on the coupling $\Omega$ but depend instead on $U$. These two states can be better appreciated in panel (b). Moreover,
at $t=0$, the state with both particles in the
lowest energy well (our initial state $|\varphi_0\rangle$) is an eigenstate with energy $E=U$, 
belonging to the central manifold.
At $t=\tau$, the state with both particles in the higher energy well
(the state with maximum charge $|\varphi_1\rangle$) is also an eigenstate with energy $E=U$, belonging to the 
central manifold as well.
\begin{figure}[t]
    \centering
    \includegraphics[width=1\columnwidth]{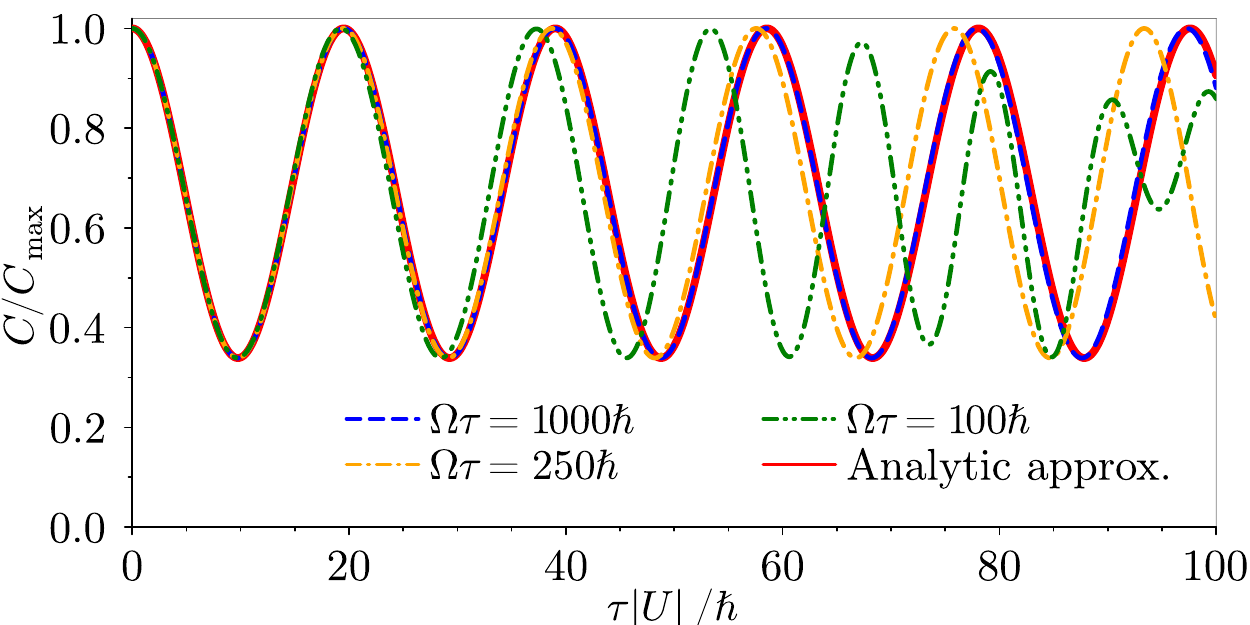}
    \caption{Charge as a function of the interaction $\tau |U|/\hbar$ for $N=2$. The solid red line shows the analytical result~(\ref{eq:analytic_charge}) for large $\Omega\tau$, while the other lines show numerical calculations for the indicated values of $\tau \Omega$.}
    \label{fig:oscillations_2}
\end{figure}

Therefore, the battery starts from a state in the central manifold, disconnected from the other states by a factor
$\Omega$. If the coupling $\Omega$ is much larger than the energy scales of the central manifold ($U$), one can expect
that all the physics will happen through the two states of Fig.~\ref{fig:spectrum_2}(b). From now on, we 
will refer to them as $|\phi_0(t)\rangle$ with energy $E_0(t)$ and $|\phi_1(t)\rangle$ with 
energy $E_1(t)$. These states are the instantaneous eigenstates of the Hamiltonian. $|\phi_0\rangle$ does not depend on time and has a constant energy $E_0(t)=U$, while  
$|\phi_1\rangle$ is time-dependent.

By computing the time evolution of the state in a superposition between $|\phi_0\rangle$ and
$|\phi_1\rangle$ with the parallel transport condition~\cite{berry_transitionless_2009}, the final ergotropy is
\begin{equation}
    C = C_\mathrm{max}\left[1-\beta\sin^2 \left(3(\pi-4)\tau U /(16\hbar)\right)\right]\,, \label{eq:analytic_charge}
\end{equation}
with $\beta= 4(2\epsilon_3-\epsilon_1-\epsilon_2)/9\epsilon_3$ [see Appendix~\ref{app:2p}]. As in Fig.~\ref{fig:charge_power}(a), Eq.~(\ref{eq:analytic_charge}) shows oscillations as a function of $\tau U$. In Fig.~\ref{fig:oscillations_2} we show a comparison between the results obtained with the numeric time evolution and the results from the analytic result~(\ref{eq:analytic_charge}). We observe an almost perfect agreement for large $\Omega\tau$, while we observe increasing discrepancies for decreasing $\Omega\tau$, as expected.
\begin{figure}[t]
    \centering
    \includegraphics[width=1\columnwidth]{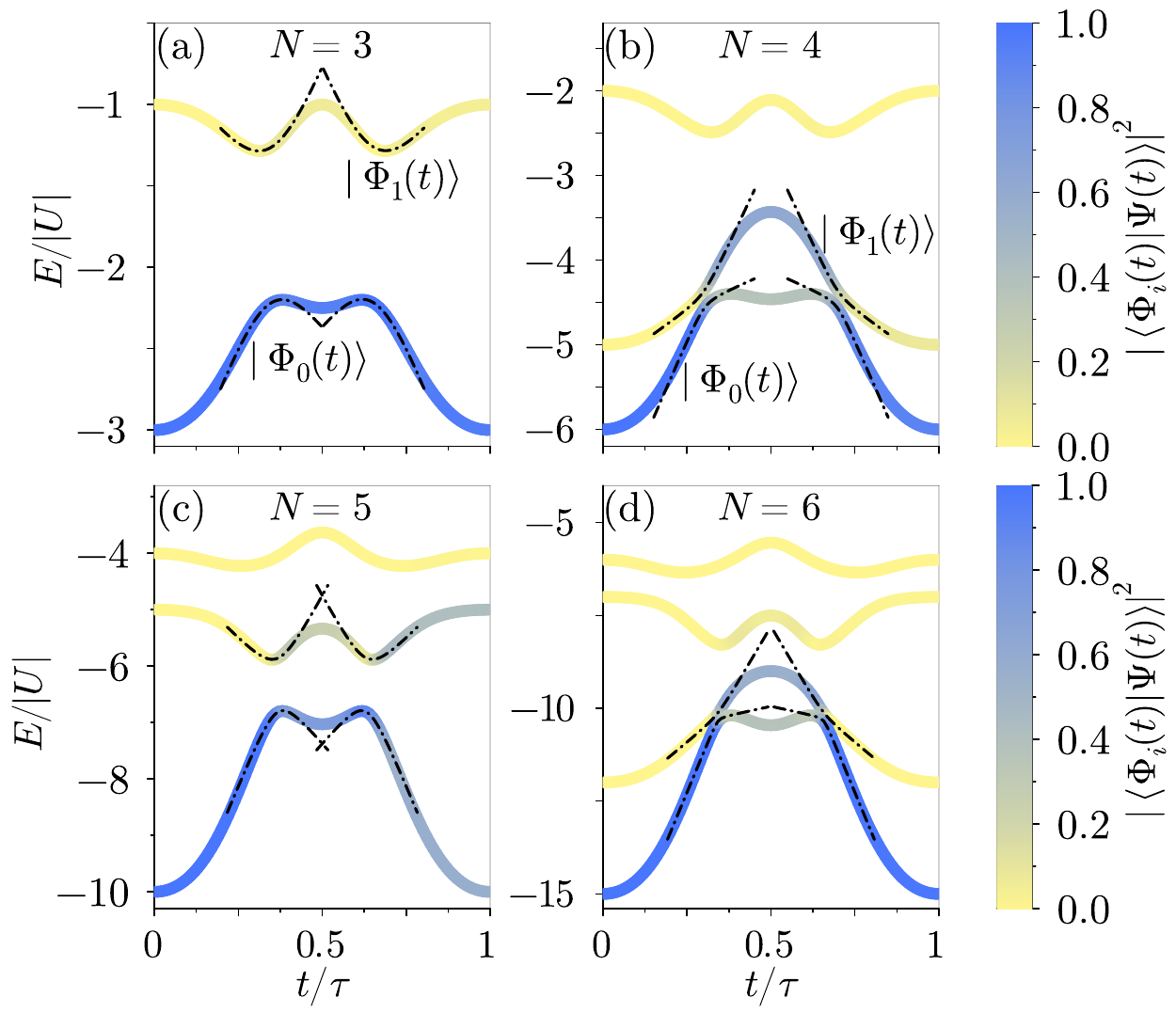}
    \caption{Energy spectrum for $N=3$ to $N=6$ and its population (in color) corresponding to a time evolution with parameters $\tau U=-20\hbar$ and $\tau\Omega=1000\hbar$. The dash-dotted lines are the fit corresponding to the energies of Eq.~(\ref{eq:LandauZener_energies}). Only shown the equivalent states of Fig.~\ref{fig:spectrum_2}(b).
    }
    \label{fig:N_spectrum}
\end{figure}

Now, we turn our attention to the general $N>2$ case, which can be analyzed with an effective theory.
Once we add more particles, the spectral properties observed with $N=2$ become slightly different. 
However, the main argument is similar; the system has a set of states with an energy
proportional to $\Omega$, whereas a few of them do not depend on that coupling. For $N=2$ only two states do not depend on $\Omega$ [Fig.~\ref{fig:spectrum_2}(b)]. However, we find that for $N$ particles  $\lfloor N/2+1\rfloor$ states fulfill such condition. 

In Fig.~\ref{fig:N_spectrum} we show these states for several numbers of particles, 
from $N=3$ to $N=6$. In contrast to Fig.~\ref{fig:spectrum_2}(b), here the states are not
degenerate at the beginning and the end of the protocol. In that case, our initial state at $t=0$
and our target state at $t=\tau$ correspond to states with energy $E/U=N(N-1)/2$, and thus, in Fig.~\ref{fig:N_spectrum}
can be identified as the manifold ground state.

In Fig.~\ref{fig:N_spectrum} the colors show the probability associated with each
instantaneous eigenstate $|\Phi_i(t)\rangle$ for a time evolution.
We obtain a slight difference between the odd and even number of particles, as we observed in the 
charge diagrams from Fig.~\ref{fig:charge_power}. For 
an odd $N$, the minimum gap between the manifold ground state (the initial state) and the first excitation of that manifold one is on the order of
$U$. On the other hand, for systems with even $N$, the minimum gap is $\sim 0.2U$. The latter indicates that for an odd $N$ it is
easier to have an adiabatic evolution due to the larger gap. For the limit of large $N$, even if the gaps still remain with the same ratios,
the relative gaps will vanish. As the total energy scales as $N^2$, the differences between 
odd and even numbers of particles will disappear in the thermodynamic limit.

For these systems, we have modeled an effective two-level model of the time-evolution. 
In that situation, the initial state (from now on, $|\Phi_0\rangle$) in Fig.~\ref{fig:N_spectrum}
has an avoided crossing with the first excitation of the manifold (from now on $|\Phi_1\rangle$) at $t\sim0.3\tau\equiv \tau_0$, and also at $t\sim0.7\tau\equiv \tau_1$. In the general case, we define $\tau_0$ and $\tau_1$ as the times where the gap is minimum.
One can assume that the evolution is adiabatic and only there is a level transition at the 
avoided crossings, with the same probabilities at both times. We also assume that all the relevant physics 
occurs between these two states, ignoring the rest.
\begin{figure}[t]
	\centering
	\includegraphics[width=\columnwidth]{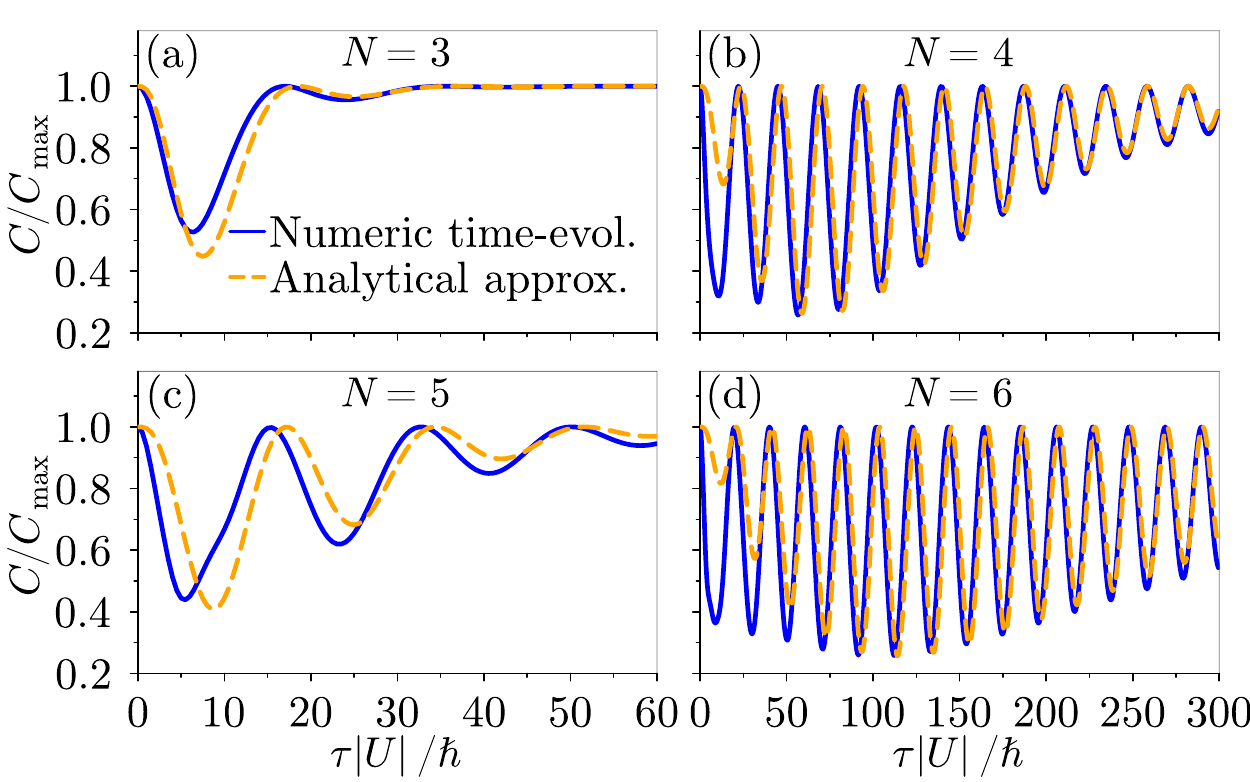}
	\caption{Charge as a function of the interaction $U$ for $N=3$ to $6$ computed with $\tau\Omega=10000\hbar$. Blue lines correspond to the numerical time-evolution calculations and the orange ones to the predictions of the model of Eq.~(\ref{eq:analytic_oscillations}), where the probability is computed using Eq.~(\ref{eq:Landau_Zener}).}
	\label{fig:oscillations_M}
\end{figure}

With that mechanism, by assuming that transitions may occur in an avoided crossing with probability $P_{\updownarrow}$, it is possible to show that the final ergotropy is [see Appendix~\ref{app:Np}]
\begin{equation}\label{eq:analytic_oscillations}
    \frac{C}{C_\mathrm{max}}=1 - 4 c \bar{P}_{\updownarrow}P_{\updownarrow}  \sin^2\left(\int_{\tau_0}^{\tau_1}(E_0-E_1)dt/(2\hbar)\right)\,,
\end{equation}
where $c = 1 - \langle \Phi_1|H_0|\Phi_1\rangle/\langle \Phi_0|H_0|\Phi_0\rangle$ and $\bar{P}_{\updownarrow}=1-P_{\updownarrow}$, and $P_{\updownarrow}$ given by the Landau-Zener transition probability,
\begin{equation}\label{eq:Landau_Zener}
    P_{\updownarrow}=\exp\left( -2\pi a^2/(\hbar|\alpha|)\right)\,,
\end{equation}
where both $a$ and $\alpha$ are obtained by the adjustment in the lowest gap in the spectrum of the two states considered. Near the transition,
in the Landau-Zener approximation, the energy of the states follows
\begin{equation}
    \label{eq:LandauZener_energies}
    E(t)=\frac{1}{2}\left((k_1+k_2)t\pm\sqrt{4a^2+(k_1-k_2)^2t^2)}\right)\,,
\end{equation}
where we fit the parameters $\alpha$, $k_1$ and $k_2$, and we compute $\alpha=k_1-k_2$.
We plot the lines of the energy obtained in the fit as dash-dotted lines in Fig.~\ref{fig:N_spectrum}.

In Fig.~\ref{fig:oscillations_M} we compare the numerical results of the final charge as a function of $\tau U$ for a large value of $\tau\Omega$ with the predictions of the two-level model using the Landau Zener 
transition probability. With our model, we only used information about the energy spectrum. We obtain a good agreement with 
the numerical time evolution simulation, especially on the frequency of the oscillations, but also in the decay of the amplitude
with more accurate results for $N=3$ and $N=4$. We observe a discrepancy for small interactions, and also
in the decay of the amplitude for large interactions, indicating that the two-level model is more robust for intermediate $U$.

\section{Summary and conclusions}\label{sec:conclusions}
In this article, we have shown how a battery based on a three-well system charges when we consider interaction. We have demonstrated that it is possible to reach the maximum charge for a finite value of the interaction through a diabatic procedure. In addition, we have explained the time evolution of the system with simple models that use only the information of the energy spectrum. 
Our quantum battery proposal is feasible experimentally in ultracold atoms laboratories nowadays.

A future perspective to extend this study could be to explore the effects of temperature and check if the battery remains stable. Additionally, it could be interesting to examine the effects of a larger chain, opening the possibility to study the effect of fermionic statistics.

\begin{acknowledgments}
We acknowledge helpful discussions with Joan Martorell and Verònica Ahufinger.  B.J-D and A.R-F acknowledge funding from Grant No.~PID2020-114626GB-I00 by MCIN/AEI/10.13039/5011 00011033 and 
"Unit of Excellence Mar\'ia de Maeztu 2020-2023” 
award to the Institute of Cosmos Sciences, Grant CEX2019-000918-M funded by MCIN/AEI/10.13039/501100011033. 
We acknowledge financial support from the Generalitat de Catalunya (Grant 2021SGR01095). 
A.R-F acknowledges funding from MIU through Grant No. FPU20/06174. F.I. acknowledges funding from ANID through FONDECYT Postdoctorado No. 3230023.
A. C. S acknowledges the support by the European Union's Horizon 2020 FET-Open project SuperQuLAN (899354), and by the Proyecto Sinérgico CAM 2020 Y2020/TCS-6545 (NanoQuCo-CM) from the Comunidad de Madrid.
NTZ acknowledges support from the Independent Research Fund Denmark and the Novo Nordisk Foundation. 
\end{acknowledgments}

\appendix
\section{Development of the two-particle time-evolution}\label{app:2p}
Our initial state $|\varphi\rangle$ can be expressed as a superposition of both states
$|\phi_0(t=0)\rangle$ and $|\phi_1(t=0)\rangle$ as 
\begin{equation}
    |\Phi(t=0)\rangle=|\varphi\rangle=\sqrt{\frac{1}{3}}|\phi_0(t=0)\rangle+\sqrt{\frac{2}{3}}|\phi_1(t=0)\rangle\,.
\end{equation}
Assuming an adiabatic evolution with the parallel transport condition,
$\langle \phi_i|\frac{\partial}{\partial t}\phi_i\rangle=0$, the
time evolution of the state can be expressed as
\begin{equation}
   |\Phi(t)\rangle= \sqrt{\frac{1}{3}}e^{-i\int E_0(t) dt/\hbar}|\phi_0(t)\rangle+\sqrt{\frac{2}{3}}e^{-i\int E_1(t) dt/\hbar}|\phi_1(t)\rangle\,,
\end{equation}
and taking into account that $E_0$ is constant, we can rearrange it as
\begin{equation}
   |\Phi(t)\rangle= \sqrt{\frac{1}{3}}|\phi_0(t)\rangle+\sqrt{\frac{2}{3}}e^{-i\int (E_1(t)-E_0) dt/\hbar}|\phi_1(t)\rangle\,.
\end{equation}
Now, by computing the integral from $t=0$ to $t=\tau$, we obtain the state at the end of the
SAP protocol reads
\begin{equation}
    |\Phi(\tau)\rangle= \sqrt{\frac{1}{3}}|\phi_0(\tau)\rangle+\sqrt{\frac{2}{3}}e^{-i3(\pi-4)\tau U/(8\hbar)}|\phi_1(\tau)\rangle\,.
\end{equation}
Note that the final state is also described by the two instantaneous time eigenstates
with a relative phase proportional to $\tau U$. 
Using this state, we can recover the charge using our definition.

\section{Development of the few particles time-evolution}\label{app:Np}
Using the two-level model, the time-dependent state will be 
\begin{equation}
  |\Psi(t<\tau_0)\rangle=|\Psi_0\rangle\,,
\end{equation}
until the first crossing. After that, there will be a transition with probability $P_{\updownarrow}$, and so, the state just after the transition will be 
\begin{equation}
  |\Psi(t=\tau_0)\rangle=\sqrt{(1-P_{\updownarrow})}|\Psi_0\rangle+\sqrt{P_{\updownarrow}}|\Psi_1\rangle\,,
\end{equation}
and with a parallel transport, we assume an adiabatic evolution until the second transition. So, the state will arrive with the phases 
\begin{align}
|\Psi(\tau_0<t<\tau_1)\rangle=&\sqrt{(1-P_{\updownarrow})}e^{-i\int_{\tau_0}^t E_0 dt/\hbar}|\Psi_0\rangle\nonumber\\
&+\sqrt{P_{\updownarrow}}e^{-i\int_{\tau_0}^t E_1 dt/\hbar}|\Psi_1\rangle\,.
\end{align}
After the second transition, we can express the state as
\begin{align}\label{eq:psi_time_tau}
|\Psi(t\geq \tau_1)\rangle=&(P_{\updownarrow}+(1-P_{\updownarrow}) e^{-i\int_{\tau_0}^{\tau_1} (E_0-E_1)dt/\hbar})|\Psi_0\rangle\\
&+\sqrt{P_{\updownarrow}(1-P_{\updownarrow})}(1-e^{-i\int_{\tau_0}^ {\tau_1} (E_0-E_1)dt/\hbar})|\Psi_1\rangle\,.\nonumber
\end{align}
Note the negative sign in the second term. This sign is needed to preserve the norm of the wavefunction, and it comes from the relative phase in the transition.
With this state, we can derive the charge at $t=\tau$.

\bibliography{QuantumBatteries}

\end{document}